\documentclass[review]{elsarticle}
\usepackage{lineno,hyperref}
\usepackage{amssymb}
\modulolinenumbers[5]
\usepackage{graphicx}
\usepackage{wasysym}

\begin{document}

\begin{frontmatter}

\title{Sensitivity of top-of-the-mountain fluorescence telescope system for astrophysical neutrino flux above 10 PeV} 
\author{A.~Neronov$^{1,2}$}
\address{$^1$Université de Paris, CNRS, Astroparticule et Cosmologie,  F-75006 Paris, France\\
$^2$Astronomy Department, University of Geneva, Ch. d'Ecogia 16, 1290, Versoix, Switzerland}

\begin{abstract}
Tau neutrinos with energies in the PeV-EeV range produce up-going extensive air showers (UEAS)  if they interact underground close enough to the surface of the Earth. This work  studies detectability of the UEAS  with a system of  fluorescence telescopes overlooking dark, low reflectivity, area on the ground up to the distances 20-50 km  from mountain top(s). Such system could provide sensitivity sufficient for  accumulation of the astrophysical neutrino signal statistics at the rate ten(s) events per year in the energy range beyond 10 PeV, thus allowing to extend the energy frontier of neutrino astronomy into 10-100 PeV range.  Comparison of sensitivities of the top-of-the-mountain telescope and  IceCube Generation II shows that the two approaches for neutrino detection are complementary, providing comparable performance in adjacent energy bands below and above 10~PeV. Sensitivity of the top-of-the-mountain fluorescence telescope system is also sufficient for the discovery of theoretically predicted cosmogenic neutrino signal. 
\end{abstract}

\end{frontmatter}

%%%%%%%%%%%%%%%%%%%%%%%%%%%
\section{Introduction}
%%%%%%%%%%%%%%%%%%%%%%%%%%%%

The astrophysical neutrino flux discovered by  IceCube in High Energy Starting Event (HESE) and through-going muon track channels \cite{icecube_discovery,IceCube_3yr,icecube_muon,icecube_icrc2017} extends as powerlaw with the slope $dN/dE\propto E^{-2.9\pm 0.3}$ \cite{icecube_icrc2017} ($2.19\pm 0.10$ in the through-going muon analysis \cite{icecube_icrc2017})  into 1-10 PeV range with no signature of a high-energy cut-off. The multi-PeV neutrinos are produced by yet unidentified cosmic particle colliders boosting protons and/or atomic nuclei energies beyond $\sim 100$~PeV. Knowledge of the highest energy properties  of the astrophysical neutrino flux is important for understanding of the mechanism of  particle acceleration in these cosmic colliders. 

The energy range above 10 PeV also should contain yet undetected ``cosmogenic'' neutrino flux generated by interactions of Ultra-High-Energy Cosmic Rays (UHECR) with cosmic microwave background and extragalactic background light during their propagation through  the intergalactic medium \cite{berezinsky,kotera,ahlers,aloisio}. This signal could come in the form of diffuse emission from sources situated at large cosmological distances, but it also could contain an identifiable "isolated source" component produced by nearby UHECR sources. 

Identification of sources and measurements of the properties of the astrophysical neutrino flux in $E\gtrsim 10$ PeV band is currently limited by the signal statistics. The IceCube detector accumulates the signal at the rate less than one event per year in the multi-PeV band \cite{icecube_discovery,IceCube_3yr,icecube_muon}. Only upper limits are derived for the flux at the energies above 10 PeV  \cite{icecube_he,auger_nu}. 

Another obstacle for the identification of neutrino sources at the highest energies  is in the limited angular and/or energy resolution of the existing neutrino detectors. The HESE event sample which is characterised by good energy resolution suffers from very large uncertainties in the reconstruction of neutrino arrival direction ($\sim 10^\circ$) \cite{IceCube_3yr}. The through-going muon neutrino sample provides good angular resolution ($\sim 1^\circ$) but suffers from limited energy reconstruction performance \citep{icecube_muon}.

Several approaches are explored for the improvement of the measurement of neutrino flux at $\gtrsim 10$ PeV. One possibility is to increase physical detection volume of ice or water Cherenkov detector, as foreseen with the IceCube Generation II \cite{icecube_gen2}, Km3NET \cite{km3net} and Baikal GVD \cite{baikal}. Better angular resolution for HESE type events is possible with water-Cherenkov (as opposed to ice-Cherenkov) detectors such as Km3NET and Baikal GVD \cite{km3net,baikal}. 

Still one more possibility is to use the Earth atmosphere rather than ice or water as part of very large volume high-energy tracker and calorimeter particle detector. The idea is to detect the flux of tau neutrinos through the up-going extensive air showers (UEAS) initiated by decaying tau leptons produced by tau neutrinos interacting below the Earth surface \cite{fargion0,fargion}.   The UEAS observations using a range of techniques of  cosmic ray physics and gamma-ray astronomy allow to track the direction of the neutrino which initiated the UEAS and provide a calorimetric measurement of the energy deposited by the neutrino interaction. The UEAS could trigger conventional surface detector arrays like those of Pierre Auger Observatory (PAO) and Telescope Array (TA) \cite{auger_nu}. They could also be observed with radio detection techniques, the idea behind TREND/GRAND project \cite{grand}. One more possibility is to use Imaging Atmospheric Cherenkov Telescopes (IACTs) similar to those used in the ground-based gamma-ray astronomy, but pointed toward the Earth surface rather than toward the sky and observing from high  altitude  as proposed for the CHANT / POEMMA, ASHRA and TRINITY telescope systems \cite{ashra,neutrino_cherenkov,chant,trinity,trinity1,venters19}.  First attempts of detection of Cherenkov light from UAES has recently been made by MAGIC telescope \cite{MAGIC}. Finally, the UEAS could be observed in fluorescence light by telescopes similar to those of PAO and TA experiments \citep{neutrino_auger_semikoz,chinese_telescope}.

This work explores the possibility to observe the UEAS with a system of fluorescence telescopes similar to (or smaller than)  those of PAO and TA, but installed at hilltops and overlooking a valley. It shows that the energy threshold of such configuration is lower than that of PAO and TA fluorescence telescopes if the monitored terrain has  low-reflectivity thus suppressing the background on top of which the UEAS are detected. Such a setup will be able to accumulate the statistics of the astrophysical neutrino signal at a rate tens of events per year, at the energies which are  factor of 10 higher than those of the highest energy astrophysical neutrinos detected so far.  It could also provide degree scale angular resolution for neutrino events together with good energy resolution (comparable to that of the IceCube HESE events)  in the energy range below  $100$~PeV, thus overcoming the limitations of the HESE and through-going muon neutrino event samples.

%%%%%%%%%%%%%%%%%%%%%%%%%%%%%%%%%%%%%%%%%%%%
\section{Tau neutrino detection through the UEAS  fluorescence signal }
%%%%%%%%%%%%%%%%%%%%%%%%%%%%%%%%%%%%%%%%%%%%

%%%%%%%%%%%%%%%%%%%%%%%%%%%%%%%%%%%%%%%%%%%%
\begin{figure}
\includegraphics[width=\linewidth]{scheme1_nu}
\caption{Principle of observations of UAES (blue) initiated by tau neutrinos (red)  with "top-of-the-mountain" telescope assemblies overlooking a valley. Showers are observed through their fluorescence and scattered Cherenkov light (yellow).  }
\label{fig:scheme1}
\end{figure}
%%%%%%%%%%%%%%%%%%%%%%%%%%%%%%%%%%%%%%%%%%%%

UAES are produced by tau neutrinos interacting below the Earth surface and producing tau leptons which decay in the atmosphere \cite{fargion0,fargion}, as illustrated in Fig. \ref{fig:scheme1}. There are different possibilities for observation of UEAS by telescopes. One option is to  place the telescope  at high enough altitude and observe a valley. An often discussed alternative is to look for showers produced by neutrinos "shooting through the mountains" \cite{fargion0,chinese_telescope}.  Furthermore, there are two possible observaiton techniques. First, it is possible to sample the fluorescence emission from air molecules excited along the shower track \cite{neutrino_auger_semikoz,chinese_telescope}. Otherwise, one could observe the Cherenkov signal from high-energy particles in the shower \cite{ashra,neutrino_cherenkov,chant}. The two approaches differ by their angular acceptance. The fluorescence emission is isotropic while the Cherenkov emission is beamed into a cone with an opening angle about $\theta_{Ch}\simeq 1.4^\circ$ (at the bottom of the Troposphere). Given comparable amount of fluorescence and Cherenkov light produced in UV band, the difference in the anisotropy of emission leads to the difference in the angular acceptance (the solid angle from which showers could be observed ($\Omega_{eff}$) and in energy threshold of the fluorescence and Cherenkov approaches. Observations in Cherenkov light allow to detect lower energy showers with small angular acceptance. Observations in fluorescence light provide large angular acceptance at the expense of increased energy threshold. 

The effective collection area of the setup consisting of a number of telescopes overlooking the ground as shown in Fig. \ref{fig:scheme1} is determined by the footprint of the telescope field-of-view (FoV) on the ground. If the telescopes are able to detect UEAS from certain direction emerging at an elevation angle $\theta$ up to a distance  is at a distance $D_{eff}$,  the collection area is  $A_{eff}=\pi D_{eff}^2\sin\theta$.  The rate of accumulation of diffuse signal from different directions is determined by the aperture, an integral of the collection area $A_{eff}$, corrected for the efficiency of neutrino-to-UEAS conversion, $p_{\nu_\tau}$, over the solid angle available for observations, $\Omega_{eff}$:
\begin{eqnarray}
\label{eq:grasp}
&&{\cal A}=\int p_{\nu_\tau}A_{eff}d\Omega\\ && \simeq 1.3\times 10^2p_{\nu_\tau}\left[\frac{D_{eff}}{20\mbox{ km}}\right]^2\left[\frac{\theta}{5^\circ}\right]\left[\frac{\Omega_{eff}}{1\mbox{ sr}}\right]\mbox{ km}^2\mbox{sr}\nonumber
\end{eqnarray}

The neutrino-to-air-shower conversion probability $p_{\nu_\tau}$  can be calculated by considering the tau  neutrino propagation through the Earth and subsequent escape of tau lepton in the atmosphere \cite{chant}. 
 It is  determined by the ratio of the maximal possible depth of neutrino interactions which result in observable UEAS, $l_\nu$, to the neutrino mean free path in the rock, $\lambda_\nu$:
$p_\nu=l_{\nu}/\lambda_\nu$. 

At the energies above $10^{15}$~eV, $l_\nu$  is estimated as the tau decay length, $\lambda_\tau\sim 5\times 10^5\left[E/10^{17}\mbox{ eV}\right]$~cm. Taking into account the neutrino charged current interaction cross-section  $\sigma_{\nu N}\simeq 4\times 10^{-33}\left[E/10^{17}\mbox{ eV}\right]^{0.3}\mbox{ cm}^2$ one finds for the estimate of $\lambda_\nu$
\begin{equation}
\lambda_\nu=\frac{m_p}{\sigma_{\nu N}\rho}\simeq 1.6\times 10^8\left[\frac{E}{10^{17}\mbox{ eV}}\right]^{-0.3}\left[\frac{\rho}{3\mbox{ g/cm}^3}\right]^{-1}\mbox{ cm,} 
\end{equation}
($\rho$ is the density of the Earth which changes between $\sim 2.6$ g/cm$^2$ and  $\sim 12$~g/cm$^3$ depending on the depth). This gives  
\begin{equation}
p_{\nu_\tau}\simeq 
3\times 10^{-3}
\left[\frac{E}{10^{17}\mbox{ eV}}\right]^{1.3}\left[\frac{\rho}{3\mbox{ g/cm}^3}\right]^{-1}
\end{equation}
in the energy range $10^{16}-10^{17}$~eV.

Tau leptons propagating through the rock loose energy via hadronic interactions on the distance scale \cite{weiler06} $l_\tau=6\times 10^5\left[E/10^{17}\mbox{ eV}\right]^{-0.2}\mbox{ cm}$,
which becomes shorter than the decay length at the energies above $\sim 10^{17}$~eV.  For neutrinos of the energies much above $10^{17}$~eV, the probability of their interaction resulting in an UEAS is almost energy-independent:
\begin{equation}
p_{\nu_\tau}\simeq
3.8\times 10^{-3}\left[\frac{E}{10^{17}\mbox{ eV}}\right]^{0.1}
\end{equation}

The mean free path of neutrinos with energies above PeV is shorter than the Earth radius $r_E\simeq 6400$~km. Because of this, Earth blocks part of neutrino flux and neutrino induced UEAS  arise at maximal elevation angle $\theta=\arcsin\left(\lambda_\nu/(2r_E)\right)\simeq7^\circ 
\left[E/10^{17}\mbox{ eV}\right]^{-0.3}$. Assuming that the showers could be arbitrarily oriented in the azimuthal direction, one finds the  effective solid angle
\begin{equation}
\Omega_{eff}=2\pi\sin\theta=\frac{\pi\lambda_\nu}{r_E}\simeq 0.8\left[\frac{E}{10^{17}\mbox{ eV}}\right]^{-0.3}\mbox{ sr}
\end{equation}
Substituting into Eq. \ref{eq:grasp} one finds
\begin{equation}
\label{eq:app}
{\cal A}\simeq 0.6\mbox{ km}^2\mbox{sr }
\left[\frac{D_{max}}{20\mbox{ km}}\right]^2
\left\{
\begin{array}{ll}
\left[\frac{E}{10^{17}\mbox{ eV}}\right]^{0.7}&,E<10^{17}\mbox{ eV}\\
\left[\frac{E}{10^{17}\mbox{ eV}}\right]^{-0.5}&,E>10^{17}\mbox{ eV}
\end{array}
\right.
\end{equation}
This dependence for fixed $D_{eff}$ is shown by the blue dotted lines in Fig. \ref{fig:aeff}.
The amount of fluorescence and scattered Cherenkov light generated by the shower is scaling with the the primary tau lepton energy, so that higher energy UEAS could be observed from larger distances. In this way, the aperture ${\cal A}$ remains a rising function of $E$ even above $10^{17}$~eV, if the orientation of the telescope is suitably chosen (not limited by the boundaries of a valley, or a mountain, not by the attenuation of the UV light). 

Convolving the aperture ${\cal A}(E)$ with the spectrum of the neutrino signal $dN_\nu/dEdt$  one could find the signal statistics expected for a given exposure time $T_{exp}$:
\begin{equation}
\label{eq:Nnu}
N_\nu=T_{exp}\int\frac{dN_\nu}{dEdt}{\cal A}(E)dE
\end{equation}
The sensitivity of the top-of-the-mountain telescope system for neutrino detection is determined by the minimal  signal statistics $N_\nu$ required for significant detection of the neutrino signal.

%%%%%%%%%%%%%%%%%%%%%%%%%%%%%%%%%%%%%%%%%%%%
\section{Estimate of the UEAS signal from Monte-Carlo simulations}
%%%%%%%%%%%%%%%%%%%%%%%%%%%%%%%%%%%%%%%%%%%%

A more detailed insight into the aperture of the top-of-the-mountain telescope system could be gained through dedicated Monte-Carlo simulations of neutrino-induced UEAS.  Such simulations include generation of isotropic neutrino  flux through sufficiently large area around the telescope site, modelling of neutrino interactions inside the Earth, propagation and decay of the tau leptons in the atmospheric volume above this area. The details of the simulation setup are given in Appendix A.

 Fig. \ref{fig:aeff} shows a comparison of the analytical formula (\ref{eq:app}) for the energy dependence of the aperture ${\cal A}$ with the result of Monte-Carlo simulations, for a system with telescopes of different diameters: $D_{tel}=1.25$~m  and $D_{tel}=2.5$~m, with optical throughput $0.85$ and photon detection efficiency of the photosensors is 0.4 in the 300-400 nm wavelength range.  These are are parameters of the IACT system of EUSO-SPB2 \cite{spb2}  and K-EUSO \cite{keuso}. The optics throughput is also the same as that of the fluorescence detectors of PAO \cite{auger}. The photon detection efficiency of the photosensors is assumed to be twice better than that of PAO (achievable e.g. with silicon photomultipliers or high quantum efficiency conventional photomultipliers). The telescope system is assumed to have the total FoV spanning  $360^\circ$ long  strip below the Earth horizon. The telescopes are installed at the altitude $H=3$~km and are overlooking a valley at the altitude close to the sea level. \textbf{We assume that the UEAS emerging from the ground at the distance $D$ are required to produce at least $50$ photoelectrons, to be detectable on top of the ground albedo background. This threshold depends on the layout of the focal surface detector and on the setup of the readout electronics (see in the next section)}.

%%%%%%%%%%%%%%%%%%%%%%%%%%%%%%%%%%%%%%%%%%%%
\begin{figure}
\includegraphics[width=\linewidth]{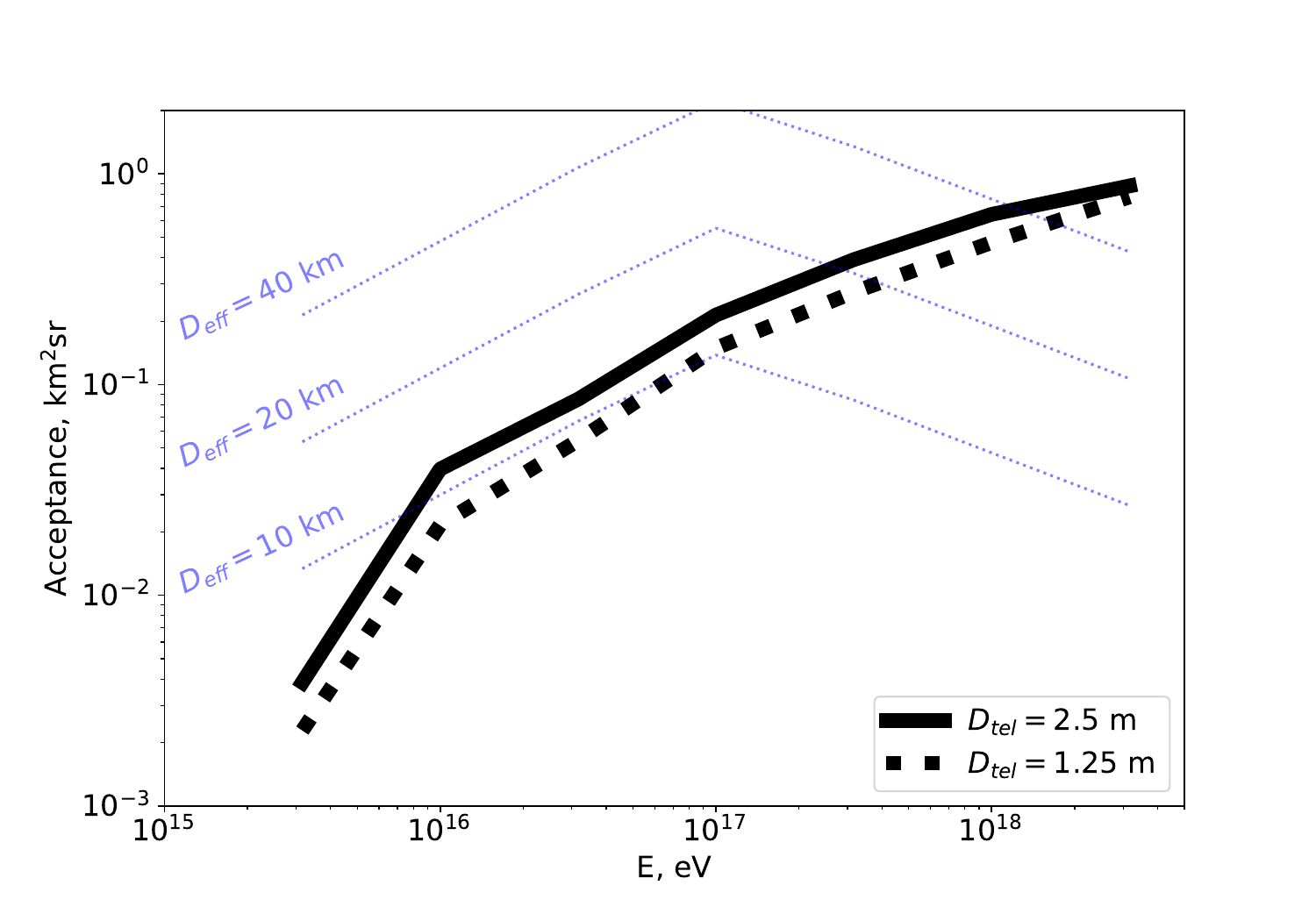}
\caption{Aperture of the system of top-of-the-mountain telescopes for detection of UAES induced by tau neutrino interactions. Dotted blue lines show a comparison with the analytical estimate of Eq. (\ref{eq:app}) for different effective maximal distances to the observable showers.}
\label{fig:aeff}
\end{figure}
%%%%%%%%%%%%%%%%%%%%%%%%%%%%%%%%%%%%%%%%%%%%

%%%%%%%%%%%%%%%%%%%%%%%%%%%%%%%%%%%%%%%%%%%%
\begin{figure}
\includegraphics[width=\linewidth]{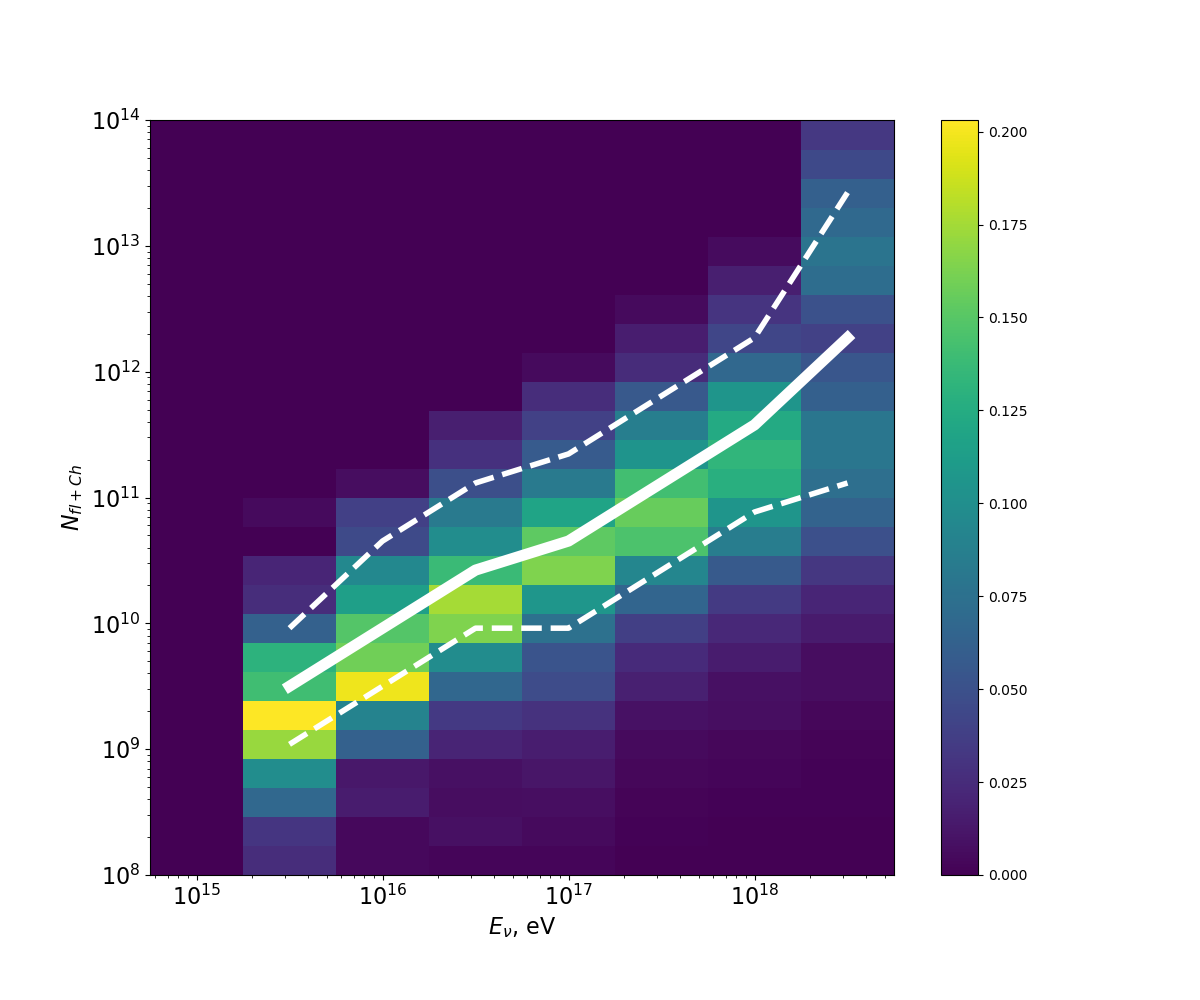}
\caption{Relation between the neutrino energy $E_\nu$ and the total fluorescence and scattered Cherenkov light luminosity (number of photons in 300-400 nm wavelength range) of the resulting UEAS. Thick solid line shows the median UEAS photon number, dashed lines show  the $\pm68\%$  containment interval around the median.}
\label{fig:E_resolution}
\end{figure}
%%%%%%%%%%%%%%%%%%%%%%%%%%%%%%%%%%%%%%%%%%%%

The numerically calculated aperture shown in Fig. \ref{fig:aeff} exhibits a sharp decrease below $10^{16}$~eV. The lowest energy of detectable UEAS is determined by the geometry of the observational setup. The span of the telescope FoV below the horizon and the shape of the mountain terrain determine minimal distance to observable showers. If the field-of-view span below the horizon is  $\Theta_{FoV}$, showers at the distances down to  $D_{min}\sim H/\tan(\Theta_{FoV})$ are observable.  The minimal distance determines the low-energy threshold of the system. Weak low energy showers should produce enough light to trigger the telescope while impacting the ground at this distance. In this respect, choosing lower altitude for the telescope $H$ location appears beneficial. However, the lowest $\sim$km height layer of the atmosphere contains an aerosol layer which scatters light and reduces the maximal distance range for high-energy shower observations. 

The  $\tau$ leptons with energies below $\sim 10^{17}$~eV decay without experiencing significant energy loss. As a result, the energy transmitted to the UEAS scales with the energy of the primary neutrino particle and could be used for the estimation of neutrino energy. Fig. \ref{fig:E_resolution} shows numerically calculated distribution of number of 300-400 nm fluorescence and scattered Cherenkov light photons $N_{fl+Ch}$ emitted by the UEAS induced by neutrinos with different energies $E_\nu$. The median UEAS luminosity, shown by the thick white line scales proportionally to the neutrino energy. The dashed lines show the 68\% containment interval around the median. They characterise the energy resolution achievable with the top-of-the mountain telescope setup. The width of the energy resolution is determined by the kinematics of the $\tau$-lepton decay in the energy range below $10^{17}$~eV. The width grows above $10^{17}$~eV is due to the additional effect of the energy loss experienced by the $\tau$ lepton before decay.

%%%%%%%%%%%%%%%%%%%%%%%%%%%%%%%%%%%%%%%%%%%%
\section{Background estimate}
%%%%%%%%%%%%%%%%%%%%%%%%%%%%%%%%%%%%%%%%%%%%

%%%%%%%%%%%%%%%%%%%%%%%%%%%%%%%%%%%%%%%%%%%%
\begin{figure*}
\includegraphics[width=\linewidth]{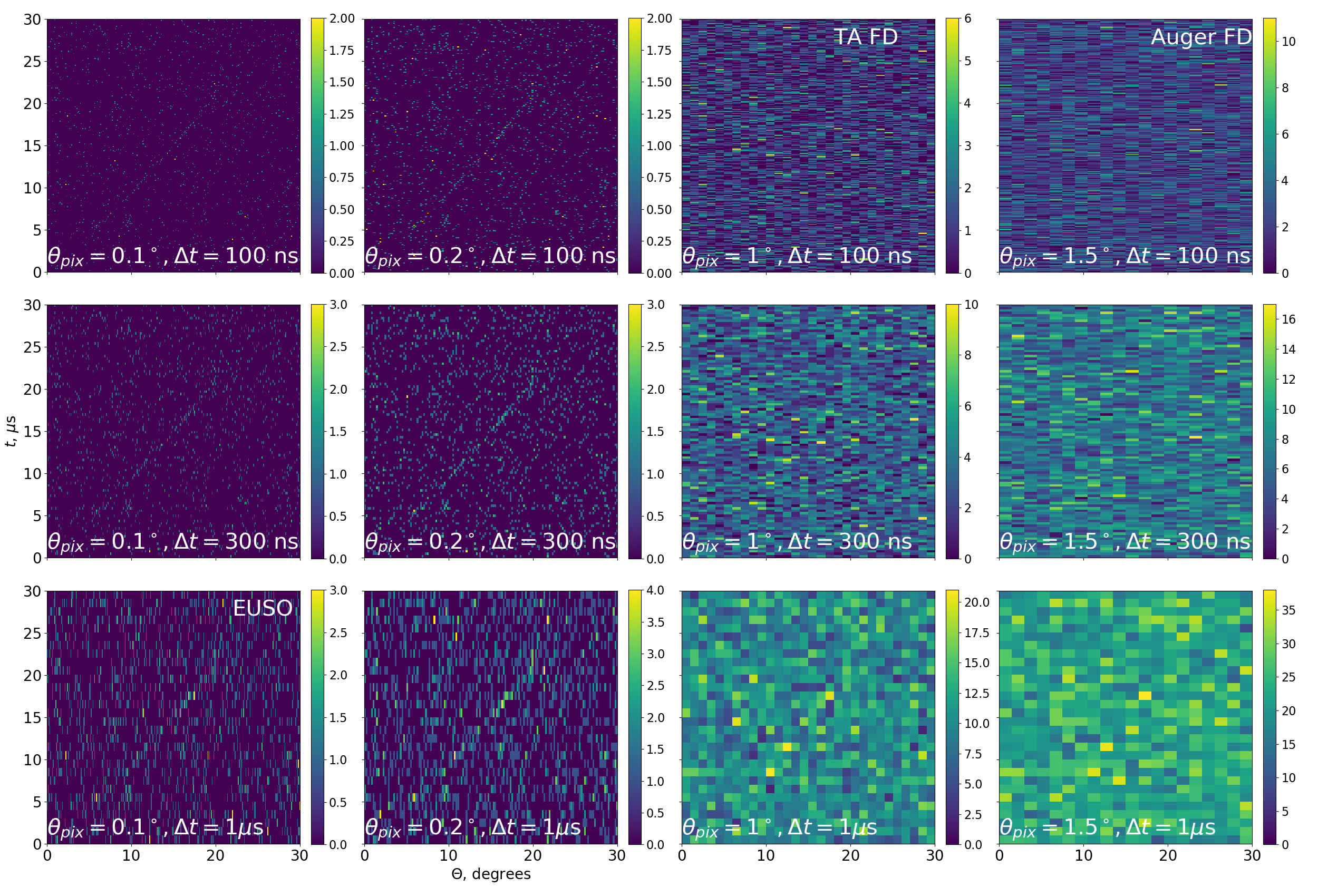}
\caption{Simulated images of an UEAS ot the size 75~photoelectrons developing at the distance 30~km from the telescope in direction normal to the line of sight. Different panels show different choices of camera pixels (columns) and sampling rates (rows). Colour shows photoelectron counts per pixel in different time bins. }
\label{fig:shower_images}
\end{figure*}
%%%%%%%%%%%%%%%%%%%%%%%%%%%%%%%%%%%%%%%%%%%%

The UEAS are detected on top of an unrelated ambient background of photons of the airglow and night sky backgrounds reflected from the ground. The reflectivity of the ground $\tilde\kappa$ is strong if it is covered by snow or ice (up to $\tilde\kappa\sim 100\%$) and is suppressed if the ground is soil (sand: $\tilde\kappa\simeq 12\%$, grass: $\tilde\kappa\simeq 8\%$, water: $\tilde\kappa<4\%$ for wavelength range $330-400$~nm \cite{uva}). The energy threshold of neutrino observations is determined by the minimal signal which is sufficiently above the level of background fluctuations. Thus, the sensitivity strongly depends on the type of terrain overlooked by the telescope.

If we consider telescope optical system with  focal distance-to-diameter ratio $F/D_{tel}\simeq 1$, the rate of the  background counts in the telescope pixels is  determined by the geometrical area of the pixels. Pixels of the size $r_{pix}=6$~mm sensitive in the 400~nm wavelength range would count at a rate \cite{euso_nadir}
\begin{equation}
{\cal R}\sim 10^{5.5}\left[\frac{\tilde\kappa}{0.1}\right]\left[\frac{\kappa}{0.3}\right]\left[\frac{r_{pix}}{6\mbox{ mm}}\right]^2\left[\frac{\lambda}{350\mbox{ nm}}\right]\mbox{ Hz}
\label{eq:backgr}
\end{equation}
where $\kappa$ is the efficiency of the optical system which includes the optics throughput and photon detection efficiency of the photosensors (assumed to be 0.85 for the optics throughput and 0.4 for the photon detection efficiency of photosensors so that $\kappa \simeq 0.3$). 

The images of nearby UAES not aligned with the line of sight span ten(s) of degrees in the camera and are accumulated on the time scale of $\sim 30\ \mu$s corresponding to the light propagation time on the distance scale $\sim 10$~km at which the shower develops. Such  signal geometry  suggests trigger and  readout scheme in which one searches for the "track-like" features in a 3-dimensional parameter space composed of camera coordinates $(x,y)$ and time $t$.  The background is minimised by an approach providing the smallest size "3D pixels" in the 3D $(x,y,t)$  space.  Reducing the size of the telescope camera pixels and choosing fine  time resolution allows to reduce the background to the minimal possible level, i.e. to collect the background which is spatially and temporally coincident with the signal in each pixel. 

\textbf{This is illustrated in Fig. \ref{fig:shower_images}, where images of showers of the size $75$ photoelectrons are shown in the position-time ($\theta-t$) coordinates for different pixel sizes and sampling rates of the telescope camera. The event is assumed to be produced by an UEAS developing at the distance 30~km perpendicularly to the line of sight.  One can see that for the assumed background rate of Eq. (\ref{eq:backgr}) the event is not detectable with large pixel ($1^\circ-1.5^\circ$) / high time resolution (10 MHz sampling rate) cameras like those of PAO \cite{auger} or TA \cite{ta}, but is detectable with small pixel / moderate time resolution  ($0.1^\circ-0.2^\circ$) camera like that of EUSO \cite{keuso,euso_nadir,spb2}, for a wide choice of sampling rates in 1-10~MHz range. }

Indeed, a  shower spanning $\sim 10^2$ pixels in the camera and spanning $t_{UAES}\sim 10\ \mu$s in time produces a signal that lasts for about $\Delta t_{pix}\sim 100$~ns in each pixel. The probability to find the signal at the level of $1$ photoelectron within this time interval in one of the pixels is $p_{pix}={\cal R}\Delta t_{pix}$. The probability to find 1 photoelectron in $n_{pix,track}\sim 10$ adjacent pixels in subsequent $\Delta t_{pix}$ time slots is $p_{pix}^{n_{pix,track}}$ (the condition of alignment of pixels along the track removes the factor $8$ for the number of different adjacent pixels to each pixel). Taking into account that the track could start in any of $N_{pix}$ pixels,  go along the direction of any other pixel, the trial factor for the arbitrary location and direction of the track is $(t_{UEAS}/\Delta t_{pix})N_{pix}$. The overall chance probability to find an $n_{pix,track}$ long track in background fluctuations is thus  $p_{camera}={\cal R}^{n_{pix,track}}\Delta t_{pix}^{n_{pix,track}-1}t_{UEAS}N_{pix}$. The number of $t_{UEAS}$ time slots within  exposure time $T_{exp}$ with duty cycle $\epsilon=0.2$ is  $\epsilon T/t_{UEAS}$, so that the overall chance coincidence probability to find a shower-like track in the camera is 
\begin{equation}
P=\frac{p_{camera}\epsilon T_{exp}}{t_{UEAS}}=\epsilon T_{exp}{\cal R}^{n_{pix,track}}\Delta t_{pix}^{n_{pix,track}-1}N_{pix}
\end{equation}
Imposing a trigger condition to have tracks of at least one photoelectron per pixel spanning at least 10-20 pixels readily reduces the rate of chance coincidence tracks from background fluctuations to the level below one per several years of observations, for a camera with as much as $N_{pix}\sim 10^5$ pixels for the reference background rate given by Eq. (\ref{eq:backgr}). 

This background estimate  shows that  already the UEAS producing tracks with several tens of photoelectron spread over 10-20 pixels  in the telescope camera are detectable with extremely low background achievable for a telescope overlooking dark terrain (rather than exposed to the night sky). Nevertheless, the statistics of 1 photoelectron per track pixel inevitably leads to gaps in the track arising due to the Poisson nature of the signal. This could be avoided by imposing higher trigger threshold at pixel level. Sensitivity estimated presented in the next section assume the threshold of 50 photoelectrons per track. 

For large enough pixels, the background grows proportionally to the size of the pixel square, because it is accumulated around the shower track in the (camera coordinates $\times$ time) 3-dimensional space.   In this respect, if we compare the fluorescence detectors of PAO, which have pixels of the angular size 1.6 degrees \cite{auger}, with those of the reference telescope system with pixel size e.g. $0.4^\circ$, one finds that the background onto which the shower signal is superimposed could be decreased by a factor $4^2=16$. This points to a possibility for suppression of background and reduction of the low-energy threshold via smaller pixel size choice. 

Apart from the reflected airglow, another type of background for neutrino detection is generated by the signal from the downward going cosmic ray EAS. These EAS produces  "Track+Cherenkov mark" type signal with the Cherenkov mark being due to the scattered Cherenkov light from the footprint of the shower on the ground. It could be efficiently rejected from neutrino exposure via timing of the "track" and "ground mark" signal and  measurement of the longitudinal profiles of the track signals.

%%%%%%%%%%%%%%%%%%%%%%%%%%%
\section{Sensitivity of the top-of-the-mountain telescope system}
%%%%%%%%%%%%%%%%%%%%%%%%%%%%

The numerically calculated aperture ${\cal A}(E)$ (Fig. \ref{fig:aeff}) could be used in Eq. (\ref{eq:Nnu}) to find the signal expected for a given neutrino flux $dN_\nu/dEdt$. For any given shape of the spectrum one could then find the normalisation which results in the signal statistics $N_\nu=1$ within a given exposure. The spectra of different models of neutrino fluxes from astronomical source populations typically result in the broad band spectra spanning several decades in energy. They could be well approximated by the powerlaw type spectra, $dN_\nu/dE=AE^{-\Gamma}$ within relatively narrow energy intervals (e.g. within one energy decade). Energy dependence of the aperture ${\cal A}$ results in the dependence of the normalisation $A_{N_\nu=1}$ of the "minimal detectable" powerlaw spectrum on the spectral slope $\Gamma$. 

\begin{figure}
\includegraphics[width=\linewidth]{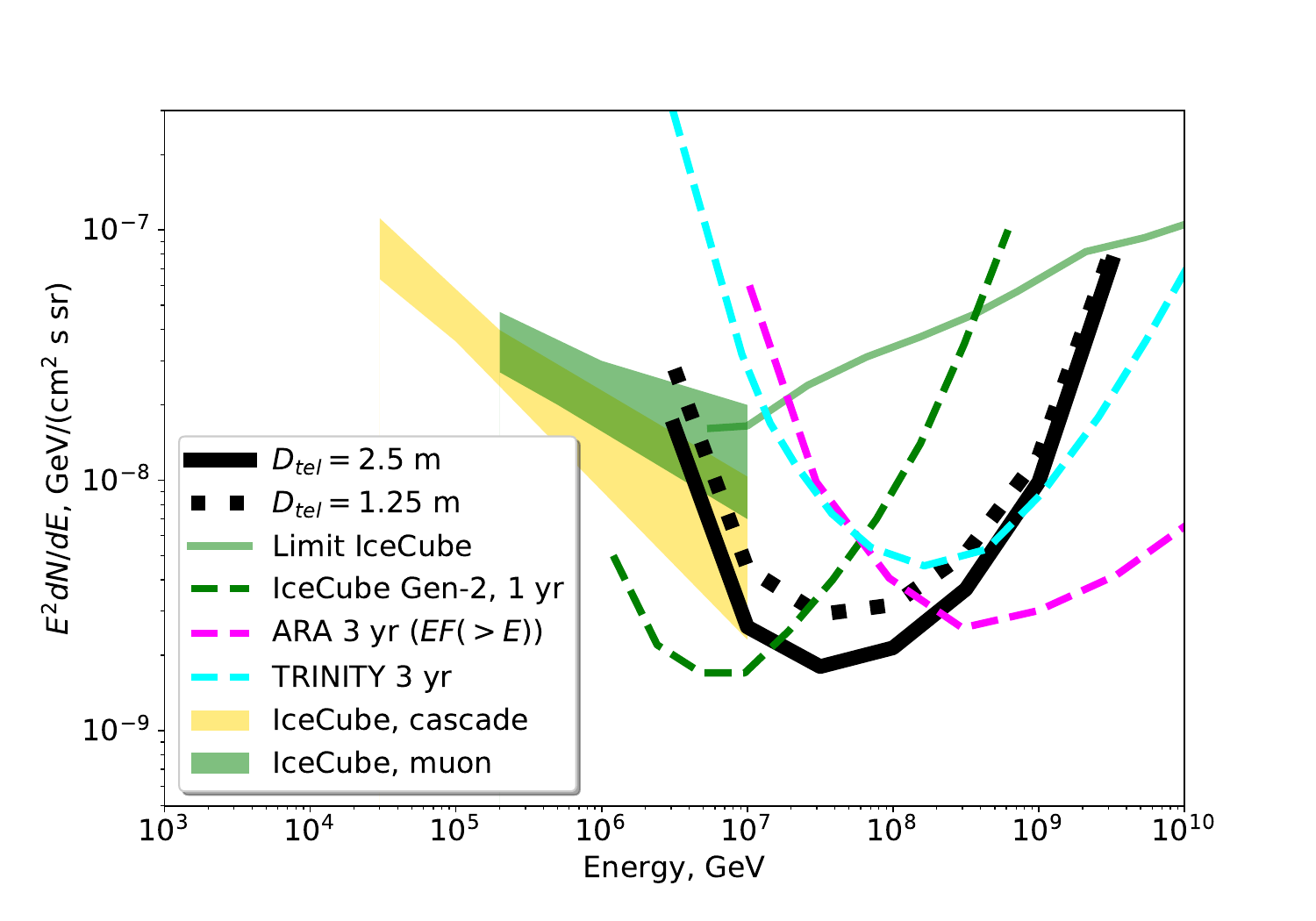}
\caption{Envelope of the minimal detectable powerlaw fluxes with the top-of-the-mountain telescope setup for year-long exposures with duty cycle 0.2. Blue data points show the IceCube signal in HESE detection mode \cite{IceCube_3yr,icecube_icrc}. Green butterfly shows the sepctrum measured in the through-going muon channel \cite{icecube_muon}. For comparison, similar envelope of minimal detectable fluxes is shown for IceCube Gen-II. The minimal detectable powerlaw type spectra are tangent to the envelope. The curves could be considered as sensitivity limit for 3-year long exposure with duty cycle 0.2. Sensitivity of TRINITY Cherenkov telescopes \cite{trinity} and ARA radio detection facility \cite{ara} are shown for comparison.}
\label{fig:sensitivity}
\end{figure}

Fig. \ref{fig:sensitivity} shows  curves which are the envelopes of the minimal detectable powerlaw spectra (those which produce $N_\nu=1$) for different values of $\Gamma$, assuming a year-long exposure with duty cycle $\epsilon=0.2$ with a telescope system overlooking $360^\circ$ strip below  the horizon km distance from a mountain top (3 km altitude). For any value of $\Gamma$ the minimal detectable powerlaw is tangent to the envelope curve. 

The envelope  curve  could also be considered as sensitivity limit for the powerlaw neutrino flux in a 3-year long exposure (in this case the powerlaw spectra which are tangent to the curve would provide event statistics $N_\nu=3$ on average and the probability to observe zero events in 3-year exposure is at the $5\%$ level). Such representation of the sensitivity also allows to judge the expected statistics of the neutrino signal for a given powerlaw type spectrum: A spectrum with normalisation $A=XA_{N_\nu=1}$ is expected to produce $X$ events within one year exposure time. This representation is also convenient for the judgement of the energy range which provides the highest signal statistics. This energy range depends on the spectral slope $\Gamma$. It could be found as the energy of the point at which the minimal detectable powerlaw spectrum touches the sensitivity curve. 

%%%%%%%%%%%%%%%%%%%%%%%%%%%%%%%%%%%%%%%%%%%%
%%%%%%%%%%%%%%%%%%%%%%%%%%%%%%%%%%%%%%%%%%%%

%From Fig. \ref{fig:sensitivity} one could see that sensitivity of the top-of-the-mountain fluorescence telescope facility is 1-2 orders of magnitude better than that of IceCube in the energy band above 10 PeV. This figure also illustrated the influence of telescope aperture on the sensitivity. Changing telescope size by a factor of 2 allows to detect weaker flux, but  induced an increase in the cost of the telescope. 

%%%%%%%%%%%%%%%%%%%%%%%%%%%%%%%%%%%%%%
\section{Discussion}
%%%%%%%%%%%%%%%%%%%%%%%%%%%%%%%%%%%%%%

The sensitivity plot in Fig. \ref{fig:sensitivity} shows that  a system of fluorescence telescopes similar or smaller than those of PAO,  overlooking the Earth surface below the horizon from a mountain top could accumulate the astrophysical neutrino flux at the rate $\sim 10$ events per year above 10 PeV  if the IceCube astrophysical neutrino spectrum extends into this energy band without a high-energy cut-off. Such a system provides a more than order-of-magnitude improvement of sensitivity compared to the IceCube at 100 PeV energy.

Observations of UEAS with fluorescence telescopes thus provide a promising approach for the study of the high-energy end of the astrophysical neutrino flux, where IceCube runs out of statistics and even IceCube Generation II telescopes would be also limited by the signal statistics.  A comparison of the sensitivity of the top-of-the-mountain setup with that of IceCube Generation II \cite{icecube_gen2} is shown in Fig. \ref{fig:sensitivity}. To estimate the sensitivity of IceCube Generation-II in the same framework as used above to the calculation of the sensitivity of the top-of-the-mountain setup, information on the expected event statistics of IceCube Generation-II events as a function of energy presented in Ref. \cite{icecube_gen2} has been used.  The event statistics for powerlaw spectra with different slopes has been recalculated (which is possible via a direct rescaling of the event statistics in each energy bin, because of the good energy resolution of  the HESE event channel). This has allowed to find  the normalisation of the spectrum which gives one event in one year exposure. The resulting green dashed sensitivity curve is the envelope of the "minimal detectable" powerlaw spectra which give one-event-per-year signal statistics. IceCube Generation II and top-of-the-mountain telescope approaches for detection of neutrino signal are obviously complementary. Combination of the data of the two facilities could provide a measurement of the astrophysical neutrino spectrum in a broad energy range up to $10^{18}$~eV. 

Comparison of the sensitivity of the top-of-the-mountain fluorescence telescope system with that of ARA radio neutrino detector ARA  \cite{ara}, also shown in FIg. \ref{fig:sensitivity}, shows that the fluorescence detection technique could reach lower energy threshold. Comparison with the sensitivity of TRINITY telescope system \cite{trinity}, which involves a similar top-of-the-mountain FoV telescope setup for detection of direct (rather than scattered) Cherenkov light shows that the fluorescence technique allows to achieve better performance in the energy range below 100 PeV. In principle, both techniques (detection of direct Cherenkov light and fluorescence $+$ scattered Cherenkov light could be combined within one and the same telescope system. This should allow to achieve further improvement of sensitivity over a wider energy range.   

%%%%%%%%%%%%%%%%%%%%%%%%%%%%%%%%%%%%%%%%%%%%
\begin{figure}
\includegraphics[width=\linewidth]{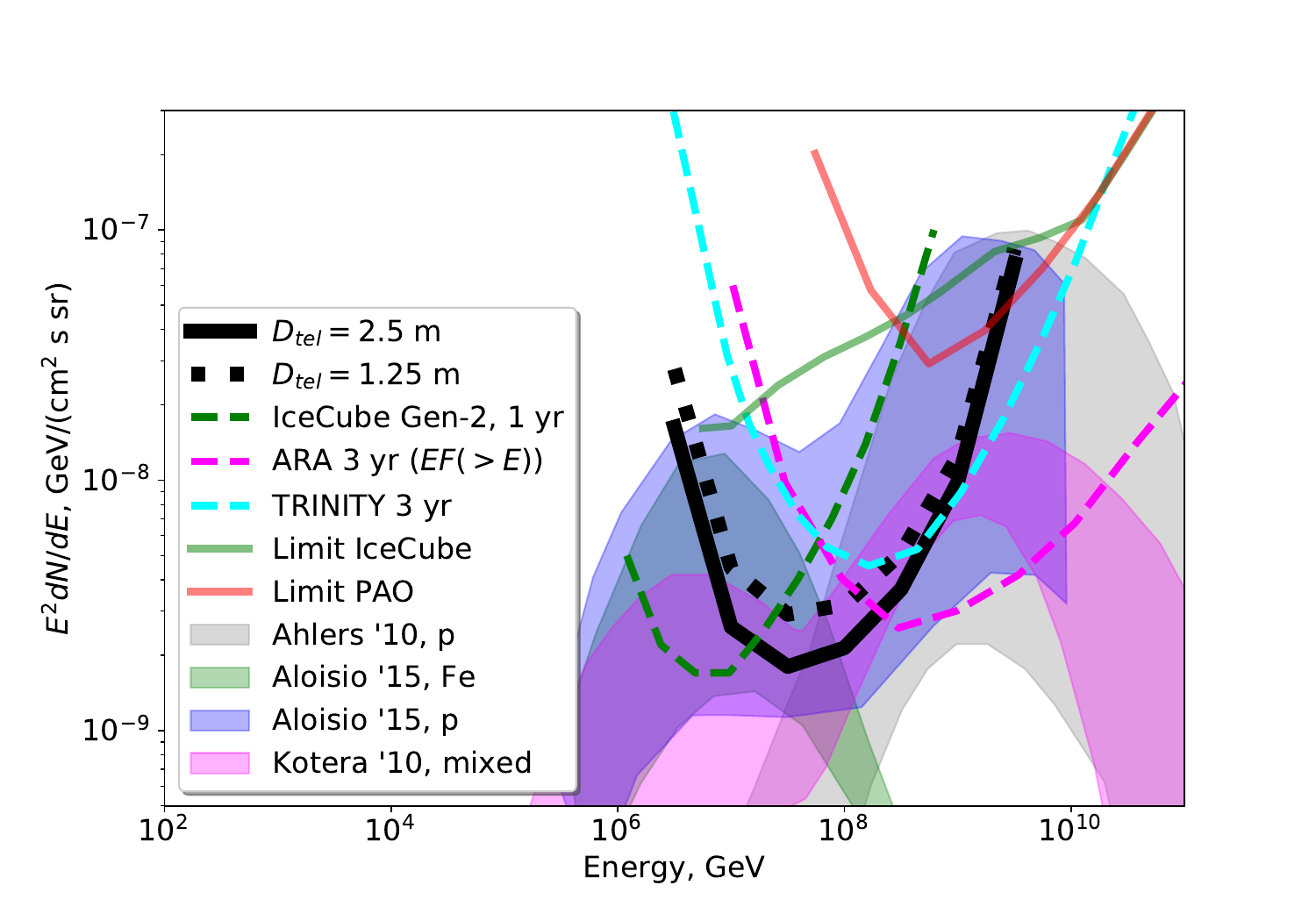}
\caption{Sensitivities of the top-of-the-mountain telescope setup compared to a range of predictions of the cosmogenic neutrino flux. Notations are the same as in Fig. \ref{fig:sensitivity}.}
\label{fig:sensitivity3}
\end{figure}
%%%%%%%%%%%%%%%%%%%%%%%%%%%%%%%%%%%%%%%%%%%%

Comparing the results with previous calculations of sensitivity of fluorescence telescopes for tau neutrino signal, we notice that in spite of similarity of the assumed telescope designs and array configurations, the setup considered above reaches significantly lower energy threshold, compared to the PAO fluorescence telescopes in neutrino detection mode, considered in  \cite{neutrino_auger_semikoz}. This is due to the exposure of the telescopes toward the ground, rather than toward the sky and due to the relative proximity and compactness of the neutrino induced UEAS with energies $10^{17}$~eV and below, compared to the cosmic ray induced EAS which were considered for energy threshold estimates in Ref.  \cite{neutrino_auger_semikoz}. 

The energy threshold of the setup considered above is consistent with that found in the Ref.  \cite{chinese_telescope} for the configuration of fluorescence telescopes looking toward a mountain surface. The setup considered above reaches higher sensitivity in roughly the same energy range because of different geometrical arrangement. The configuration in which the telescopes overlook a mountain from a valley has fixed geometrical collection area determined by the footprint of the telescope field of view on the mountain. To the contrary, the configuration in which telescopes overlook the valley from the mountain top allows to achieve effective area which grows with energy, because higher energy showers are observable from larger distances. 

Observations of the UEAS through the fluorescence emission provide a calorimetric measurements of the energy of  the shower, because the fluorescence yield is directly proportional to the energy deposition of the shower particles in the air. With sufficient signal statistics of $S \gtrsim 100$ photoelectrons, the energy resolution of such measurements could be as good as $\lesssim S^{-1/2}\sim 10\%$ (without account of systematic effects). Tau leptons with energies below $10^{17}$~eV decay before they loose energy on interactions with the medium. Because of this, the fluorescence detection modes could serve as a proxy for the $\tau$ lepton and hence to the primary neutrino energy  in the energy range below $10^{17}$~eV, as illustrated by Fig. \ref{fig:E_resolution}. In this energy range the energy resolution of the top-of-the-mountain fluorescence telescope  system  (half-decade in energy) is mostly limited by the kinematics of the $\tau$ lepton decay. At higher energies, tau leptons suffer from energy loss and emerge from the ground with energy reduced by unknown fraction. This limits the energy resolution. The situation in this case is similar to the case of neutrino detection in the through-going muon track channel in IceCube. Information on the primary neutrino spectrum could be obtained only via statistical analysis of a large number of events. 

The top-of-the-mountain setup allows also to achieve good angular resolution $\lesssim 1^\circ$. The direction of a  shower image spanning 50 pixels in the camera could be reconstructed with down to $\delta\theta=1/50\simeq 1^\circ$ precision, which is comparable to the angular resolution of the water / ice Cherenkov detectors IceCube and Km3NET in the through-going muon track detection channels \cite{icecube_gen2,km3net}.

Fig. \ref{fig:sensitivity3} shows that the top-of-the-mountain telescope exploration of the energy band $>10$~PeV will provide a one-two order of magnitude improvement of sensitivity with respect to the existing limits from IceCube  \cite{icecube_he} and PAO \cite{auger_nu}. This improvement will allow to explore a range of models of cosmogenic neutrino flux \cite{ahlers,kotera,aloisio} for different UHECR flux composition models and different assumptions about high-energy cut-offs in the source spectra and cosmological source evolution.  

It is useful to note that the top-of-the-mountain fluorescence telescopes could be used not only for the UEAS detection, but also for the detection of conventional downgoing EAS induced by cosmic rays, if the telescopes are tilted to higher elevation angles. The neutrino and cosmic ray observation modes are in fact complementary. Tracing of down-going cosmic ray EAS is possible at good weather conditions in the absence of clouds. To the contrary, the UEAS produced by neutrinos develop low in the Troposphere. Presence of clouds in the altitude range above $\sim 3-5$~km does not prevent observations of the low-altitude UEAS. Combination of neutrino and cosmic ray observation modes thus provides a possibility to increase the duty cycle of operation of the fluorescence telescopes.  

%%%%%%%%%%%%%%%%%%%%%%%%%%%%%%%%%%%%%%%%%
\appendix
\section{Monte-Carlo simulations of the UEAS}
%%%%%%%%%%%%%%%%%%%%%%%%%%%%%%%%%%%%%%%%%

Estimates of the sensitivity of the top-of-the-mountain telescope system for tau neutrino deteciton is done based on the Monte-Carlo simulation of neutrino-induced UEAS. This appendix provides information on the details of the simulation setup. 

The simulations start from generation of the neutrino trajectories that are required to pass through a horizontal disk of the radius $\sim 50$~km around the telescope site. The directions of neutrino trajectories are homogeneously distributed over a half-sphere facing upward (positive elevation angles). For each neutrino energy bin between $10^{15}$~eV and $3\times 10^{18}$~eV, some $10^7$ neutrino trajectories are simulated. 

At the second step for each neutrino trajectory an interaction distance $d_\nu$ is drawn from an exponential distribution with the width equal to the energy-dependent mean free path $\lambda_\nu(E_\nu)$, taking into account the depth dependence of the Earth density. This distance is compared to the path length through the Earth. If the interaction distance is longer than the path through the Earth,  modelling of the neutrino event stops and next neutrino trajectory is generated. 

Otherwise, the modelling proceeds with the $\tau$ lepton generated in the neutrino interaction. The initial energy of the tau lepton, $E_{\tau,0}$ is lower than the energy of the neutrino, $E_\nu$, because a part of energy is taken by the hadronic shower at neutrino interaction point. We use the result reported in Ref. \cite{thesis} on the distribution of inelasticity $\kappa$ in the charged current tau neutrino interaction to draw $E_{\tau,0}$ from the distribution  $(1-\kappa)E_\nu$. 

The energy loss rate of tau is 
$dE/dX=\rho \beta E$; $\rho$ is the density of the rock and $\beta\simeq 5\times 10^{-7}$cm$^2$/g in the energy range of interest \cite{tau_loss,jeoung17}. This imples that energy loss distance $l=1/(\beta\rho)\simeq 7$ km and tau energy changes as  $E(x)=E_0\exp(-x/l)$, where $E_0$ is its energy at production. Taking into account that tau decays on the distance scale $\lambda_\tau$, one could find tau lepton survival probability as a function of distance $x$ (see e.g. \cite{thesis})
$$P(x,E_0)\propto \exp\left(-\frac{m_\tau}{\tau_\tau}\int\frac{dx}{E(x)}\right)=\exp\left(-\frac{m_\tau l}{\tau_\tau E_{0,\tau}}\exp\left(\frac{x}{l}\right)\right)$$. 
The tau lepton produced in neutrino interaction is supposed to follow the same direction as the neutrino and its propagation distance $l$ before decay is drawn from the distribution $P(x,E_0)$. 
If $l+d_\nu<D$, the calculation stops and next neutrino trajectory is generated.  The analysis does not take into account the neutrino regeneration effect which could significantly indcrease the tau neutrino to UEAS conversion probability at the energies above $10^{18}$~eV \cite{alvarez19}.

Otherwise, if $d_\nu<D$ and $d_\nu+l>D$, the tau lepton decays in the atmosphere and produced an UEAS. The calculation of the UEAS initiated by $\tau$ leptons uses CORSIKA EAS simulation program \cite{corsika}.   The fluorescence yield of the UEAS is determined by the energy dissipation rate at every step of the longitudinal profile. The fluorescence yield in the Troposphere is assumed to be 20 photons per MeV energy dissipation.  The longitudinal profiles of the charged particle content of the showers are used to calculate the longitudinal profile of the  both fluorescence and Cherenkov light (which also contributes to the UEAS signal). 

The fluorescence and Cherenkov photons are propagated  through the atmosphere taking into account attenuation of the signal by the Rayleigh scattering in the atmosphere.   This effect is characterised by the optical depth  
\begin{equation}
\tau\simeq 0.7\left[\frac{\lambda}{350\mbox{ nm}}\right]^{-4}\left[\frac{R}{10\mbox{ km}}\right]
\end{equation}
where $\lambda$ is the wavelength.   
the atmosphere is not transparent for the UV and blue light in the wavelength range $\lambda\lesssim 400$~nm already on the distance range $D\sim 20-30$~km.  In addition, attenuation due to aerosols is taken into account. The aerosols are assumed to be distributed close to the ground with the vertical optical depth $0.05(\lambda/500\mbox{ nm})^{-1}$ and the scale height $H_a=1$~km. 

The observation point is assumed to be situated at the altitude $H_{obs}=3$~km overlooking a $360^\circ$ strip below the Earth horizon up to the maximal distance to the generated neutrino trajectories (taken to be $50$~km). The UEAS emerging from the ground  at the distance $5$~km from the telescope are assumed to trigger telescope is they generate a signal stronger than $N_{thr}=50$~photoelectrons. This assumption is based to the estimate of the background levels presented in the main part of the paper.

The full sample of the simulated neutrino induced UEAS events is finally used to calculate the energy dependent acceptance  ${\cal A}$ shown in Fig. \ref{fig:aeff}.  For this calculation  the geometrical acceptance of a hypothetical 100\% efficient UEAS detector sampling the UEAS from the full 50~km radius region around the telescope is multiplied by the fraction of events detectable above the threshold, weighted by the sine of the elevation angle (see Eq. \ref{eq:grasp}).   

\section*{Acknowledgement}

This work was partially supported by the Ministry of science and higher education of Russian Federation under the 
contract 075-15-2020-778 in the framework of the Large scientific projects program within the national project "Science".

\bibliography{UHECR_Cherenkov,NUCRA_neutrinos}

\end{document}